\documentclass[epj]{svjour}
\usepackage{graphics}
\usepackage{xcolor}
\usepackage{float}

\begin{document}
\title{Neutron transfer in $^{9}$Be + $^{159}$Tb system}

\author{Malika~Kaushik\inst{1}, G.~Gupta\inst{2}, V.V.~Parkar\inst{3,4},  S.K.~Pandit\inst{3}, Swati~Thakur\inst{1}, V.~Nanal\inst{2} \thanks{\emph{email:nanal@tifr.res.in} }, A.~Shrivastava\inst{3,4}, R.G.~Pillay\inst{1}, H.~Krishnamoorthy\inst{4,5}, K.~Mahata\inst{3,4}, S.~Pal\inst{6}, C.S.~Palshetkar\inst{2},  K.~Ramachandran\inst{3}, and Pushpendra~P.~Singh\inst{1}}                     
%
%
\institute{Department of Physics, Indian Institute of Technology Ropar, Rupnagar - 140 001, Punjab, India\and  Department of Nuclear $\&$ Atomic Physics, Tata Institute of Fundamental Research, Mumbai - 400005, India\and  Nuclear Physics Divison, Bhabha Atomic Research Centre, Mumbai - 400085, India\and  Homi Bhabha National Institute, Anushaktinagar, Mumbai - 400094, India \and  India-based Neutrino Observatory, Tata Institute of Fundamental Research, Mumbai - 400005, India \and Pelletron Linac Facility, Tata Institute of Fundamental Research, Mumbai - 400005, India}

\date{Received: date / Revised version: date}
%
\abstract{
One neutron stripping cross sections ($\sigma_{-1n}$) are measured in $^{9}$Be+$^{159}$Tb system in the energy range  E$_{cm}$/V$_{B}$ $\sim$ 0.79 - 1.24 using offline gamma counting technique. The CRC model calculations including the ground state and the 2$^{+}$ resonance state of $^{8}$Be, carried out using the FRESCO code, give a reasonable description of the  measured data. In addition, comparisons of reduced 1n-stripping cross sections- $\sigma_{red}$ with $^{9}$Be for  different target nuclei (A $\sim$150-200),
and $\sigma_{red}$ for $^{9}$Be, $^{6}$Li with $^{159}$Tb target are presented. While no strong target dependence is observed with $^{9}$Be projectile, $\sigma_{red}$($^{9}$Be) is significantly larger than that for $^{6}$Li, which is consistent with the Q-value for transfer reactions and breakup threshold energy of projectiles.
}
%
\titlerunning{Neutron transfer in $^{9}$Be + $^{159}$Tb system}
\authorrunning{Malika Kaushik et al.}
\maketitle
\section{Introduction}
\label{intro}
The reactions involving weakly bound nuclei at energies around the barrier have been extensively studied in recent years. While the major focus has been on the study of the complete and incomplete/breakup fusion (ICF/BUF) mechanisms~\cite{vj,lf}, transfer reactions have also attracted significant attention~\cite{skp-PLB2021,Lei_MoroPRL}.
 The transfer reactions, particularly with weakly bound projectiles, are useful to probe the role of valence nucleons~\cite{as,al,yep,ap,an} and can provide important insight into reaction dynamics and nuclear structure aspects~\cite{skp-PRC2019,vvp,skp,sph,ydf}.
  With weakly bound projectiles (e.g. $^{6,7}$Li, $^{9}$Be), low breakup threshold plays an important role in reaction dynamics. Additionally, for $^{9}$Be, the  neutron separation energy is relatively low (S$_{n}$ = 1.66 MeV~\cite{nndc}). Consequently, 1$n$-stripping, the process in which a valence neutron transfers from projectile to the target nucleus, has a significant contribution to the total reaction  cross section at subbarrier energies ~\cite{al,ap,adp}.
  
Recent studies with $^{9}$Be, have shown  that $^{8}$Be~+~$n$ is a dominant configuration ~\cite{vvp,skp,tad}, as compared to $\alpha$ + $\alpha$ + $n$ or $^{5}$He~+~$\alpha$~\cite{nk}. 
Several experiments have probed impact of 1$n$- stripping on other reaction channels~\cite{vvp,skp,sph,ydf1}. 
Experimentally, direct one-step process (1$n$ transfer) or a two-step process (incomplete fusion)- where $^{9}$Be breaks into $^{8}$Be~+~$n$ ($E_{\rm BU}$ $\sim$ 1.66 MeV) and then the neutron fuses with the target,  can not be distinguished in inclusive measurements.
This is also illustrated in $^{9}$Be~+~$^{186}$W system~\cite{ydf1}, where a comparison  of $\sigma$(fusion + $n$-transfer) with  universal fusion function (UFF) at E~=~44~MeV ($>V_{B}$)  showed no effect of suppression. 
Systematic exclusive measurements of breakup in $^{9}$Be reactions over a wide range of target nuclei $^{144}$Sm to $^{209}$Bi at near barrier energies have shown that the probability of breakup is nearly independent of the target nucleus ~\cite{rr}.
Although $1n$-stripping is a major contributor to the breakup process, target dependence of neutron transfer process can be separately probed in  $^{9}$Be-induced transfer reactions.

Recently,  it is shown that transfer cross sections on the $^{197}$Au target with $^{9}$Be are considerably higher than those with  $^{6,7}$Li~\cite{mk1}. 
Transfer reaction studies with weakly bound projectiles on $^{159}$Tb target are rather sparse. Influence of the $n$-transfer cross sections in $^{6}$Li~+~$^{159}$Tb reaction was studied by Pradhan {\it et al.}~\cite{mkp}. 
The comparison of the measured cross section with DWBA (Distorted Wave Born Approximation) calculations indicated spectroscopic factor of 0.25 for 63.68 keV state of $^{160}$Tb.
As the transfer process largely depends on projectile-target combination, it would be interesting to compare the 1$n$-stripping  cross section for the $^{159}$Tb target with different weakly bound stable projectiles, namely,  $^{6}$Li and $^9$Be.\\

With this motivation, 1$n$-stripping cross section in $^{9}$Be + $^{159}$Tb system from sub to above barrier energy range is reported in this paper. A brief description of  experimental setup and data analysis is given in the second section. The theoretical analysis of the measured 1$n$-transfer cross sections together with systematics for $^{9}$Be~+~X  are presented in section~3, followed by a summary section.

\section{Experimental Details and Data Analysis}
The experiment of $^{9}$Be + $^{159}$Tb system was performed using offline gamma counting technique at BARC-TIFR Pelletron Linac Facility, TIFR, Mumbai, India. The details of experimental setup are described in Ref.~\cite{mk}. The $^{9}$Be beam (E = 30 - 47 MeV) was bombarded onto the self-supporting $^{159}$Tb  target ($\sim$ 1.3-1.7 mg/cm$^{2}$) + Aluminum  catcher foil (1.5 mg/cm$^{2}$)assemblies. Either (Tb-Al)-(Tb-Al) or (Au-Al)-(Tb-Al) stack arrangement was deployed at different beam energies for the optimal use of the beam time.  The energy loss in foils, and consequently the incident beam energy at the center of the target were calculated  using  SRIM~\cite{sr}. Table~\ref{tab:1} gives details of target stacks and beam energies. The offline gamma counting was carried out using an efficiency calibrated HPGe detector, in a close geometry (i.e. with the irradiated foil mounted on the face of the detector). Since $^{160}$Tb has relatively long half-life ($T_{1/2}\sim$ 72.3 d)~\cite{nndc}, the offline counting was typically done after six to ten days of irradiation, which helped to reduce the background.

\begin{figure}[h]

\resizebox{0.5\textwidth}{!}{%
  \includegraphics{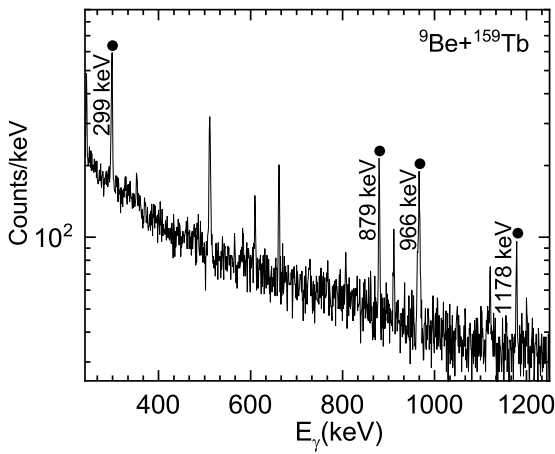}
}
\caption{A typical $\gamma$-ray spectrum of $^{9}$Be+$^{159}$Tb system at E$_{lab}$= 37 MeV, recorded after a cooldown period of 8.5 d (t$_{collection}$~=~11.9 h). Some of the dominant characteristic $\gamma$ rays of $^{160}$Tb (corresponding to 1$n$-stripping  channel)  are marked.}
\label{fig:1}       
\end{figure}

\begin{figure}
\resizebox{0.5\textwidth}{!}{%
  \includegraphics{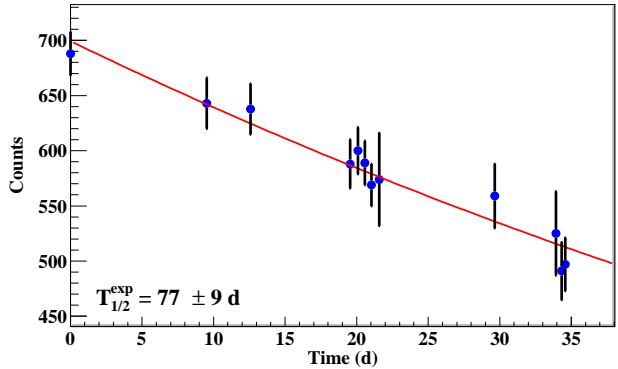}
}
\caption{Measured decay curve of 879 keV (characteristic gamma ray of $^{160}$Tb).}
\label{1t}       
\end{figure}
Fig.~\ref{fig:1} shows a typical offline $\gamma$ ray spectrum of $^{9}$Be~+ $^{159}$Tb system at E$_{lab}$ = 37 MeV, where  the dominant characteristic $\gamma$-rays of $^{160}$Tb,  corresponding to 1$n$-stripping channel, are marked. For further analysis 879 keV $\gamma$ ray ($I_\gamma= 30\%)$, which was cleanly visible at all energies was used.
 The half-life of 879 keV was measured to be 77$\pm$ 9 days, can be seen in Fig.~\ref{1t}, which is consistent with the reference value~\cite{nndc}. The cross sections of $^{160}$Tb have been calculated following the analysis procedure described in Ref.~\cite{mk} and are given in Table~\ref{tab:2}.

\begin{table}
\centering
\caption{Details of target-catcher foil sets along with the incident beam energy incident (E$_{inc}$ (MeV)).  E$_1$ and $E_2$ refer to lab energies at the centre  of the target of Set 1 and Set 2, respectively. }
\label{tab:1}       
\begin{tabular}{ccccc}
\hline\noalign{\smallskip}
E$_{inc}$  & E$_1$  & Set 1 & E$_2$  & Set 2 \\
(MeV) & (MeV) &  &  (MeV) & \\
\noalign{\smallskip}\hline\noalign{\smallskip}
30.5 & 30.1 & Tb+Al & 27.1  & Tb+Al\\
33.5 & 33.1 & Tb+Al & 30.3 &Tb+Al \\
37.5 & 37.0 &Tb+Al &- &- \\
41.0 & 40.6 &Au+Al & 38.4 & Tb+Al \\
45.0 & 44.7 &Au+Al & 42.6 &Tb+Al\\
\noalign{\smallskip}\hline
\end{tabular}
\end{table}

\begin{table}
\centering
\caption{Measured cross sections of 1$n$-stripping channel in $^{9}$Be~+~$^{159}$Tb system (V$_{B}$ = 32.5 MeV). Errors are statistical (inclusive of fitting errors).}
\label{tab:2}       
\begin{tabular}{lll}
\hline\noalign{\smallskip}
E$_{lab}$ (MeV) & E$_{cm}$ (MeV)& $^{160}$Tb(mb)  \\
\noalign{\smallskip}\hline\noalign{\smallskip}
27.1 & 25.6 & 13 $\pm$ 1 \\
30.1 & 28.5 & 30 $\pm$ 2 \\
30.3 & 28.7 & 33 $\pm$ 4 \\
33.1 & 31.3 & 83 $\pm$ 12 \\
37.0 & 35.0 & 138 $\pm$ 16 \\
38.4 & 36.3 & 168  $\pm$ 20 \\
42.6 & 40.3 & 175 $\pm$ 20 \\
\noalign{\smallskip}\hline
\end{tabular}
\end{table}
\begin{table*}

\caption{Potential parameters used in CRC calculations for $^{9}$Be, $^{6}$Li~+~$^{159}$Tb system. The $R_{i} = r_{i}.A^{1/3}$, where \textit{i~=~R, V, S, C} and \textit{A} is the target mass number.}
\label{tab:3}       
\begin{tabular}{lllllllllll}
\hline\noalign{\smallskip}
 System &V$_{R}$ (MeV) &  r$_{R}$ (fm) & a$_{R}$ (fm) & W$_{V}$(MeV) & r$_{V}$(fm) &a$_{V}$(fm) &  W$_{S}$(MeV) &r$_{S}$(fm)  & a$_{S}$(fm) &  r$_{C}$ (fm) \\ \hline

\noalign{\smallskip}\hline\noalign{\smallskip}
$^{9}$Be~+~$^{159}$Tb~\cite{yl} & 258.80    & 1.35   & 0.73 & 15.15   & 1.64 & 0.60 & 46.82  & 1.20 & 0.84  & 1.56\\

 $n$~+~$^{8}$Be~\cite{jl} & 50.00  \footnotemark[1]  & 1.15   & 0.57 & - & - & - & 5.50  & 1.15 & 0.57  & -\\

$n$~+~$^{159}$Tb~\cite{dgk} & 50.00  \footnotemark[1]   & 1.23   & 0.65 & -   &- & - & 6.00  & 1.23 & 0.65  & -\\

$^{6}$Li~+~$^{159}$Tb~\cite{jc} & 109.5  & 1.33 & 0.81 & 24.97 & 1.53 & 0.88 & &  &  & 1.3\\

 $n$~+~$^{5}$Li~\cite{jc} & 50.00  \footnotemark[1]  & 1.25   & 0.70 & - & - & - & 6.0  & 1.25 & 0.70  & -\\
\noalign{\smallskip}\hline

\end{tabular}

\footnotemark{Depth adjusted to obtain the correct binding energy.}
\end{table*}

\section{CRC Calculations}

 The measured excitation functions of the 1$n$-stripping channel ($^{160}$Tb) are analysed in the framework of coupled reaction channel (CRC) calculations using theoretical model code FRESCO~\cite{ijt}.  The Woods-Saxon form is employed for both real and  imaginary parts of the optical potential. The potential parameters employed in calculations are tabulated in Table~\ref{tab:3}. The global optical potential parameters for $^{9}$Be~+~$^{159}$Tb are taken from Ref.~\cite{yl}. 
The binding potential parameters between the transferred nucleon and the core, $\textit{i.e.,}$ for $n$~+~$^{8}$Be are taken from Ref.~\cite{jl}, while those for $n$~+~$^{159}$Tb are taken to be same as that for $n$~+~$^{208}$Pb~\cite{dgk}. 
  The calculations are done including the energy states upto 1.4~MeV for $^{160}$Tb which are listed in Table~\ref{tab:4}, with the spectroscopic factors (C$^{2}$S) of all the target states 
  as 1.0. For $^{8}$Be, the ground state  with C$^{2}$S =0.42~\cite{jl} and the 2$^{+}$ resonance state (E = 3.03~MeV) with C$^{2}$S =1.0 are taken into consideration.  

\begin{figure}[h]

\resizebox{0.5\textwidth}{!}{%
  \includegraphics{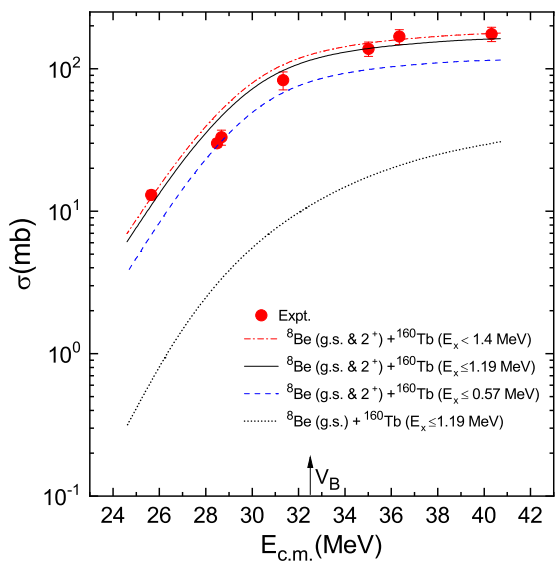}
}
\caption{Measured 1$n$-stripping excitation function ($^{160}$Tb) together  with  CRC calculations. The ground state and 2$^{+}$ resonance (E = 3.03 MeV) state of $^{8}$Be and different combinations of excited states of $^{160}$Tb are considered in the calculations.}
\label{fig:3}       
\end{figure}

\begin{table}[H]
\centering
\caption{Excited states of $^{160}$Tb~\cite{nndc} included in CRC calculations (C$^2$S=1). Only confirmed $J^\pi$ values are listed. }
\label{tab:4}       
\begin{tabular}{llllll}
\hline\noalign{\smallskip}

 E (MeV) & J$^{\pi}$ & E (MeV) & J$^{\pi}$ & E (MeV) \\ [0.5ex] 
 \hline
0.0 & 3$^{-}$ &0.269 & 2$^{+}$ & 0.599    \\

 0.064 & 1$^{-}$ & 0.279 & 4$^{-}$ & 0.660        \\
 0.064 & 4$^{+}$ & 0.318 & 3$^{+}$&  0.684    \\

0.079 & 4$^{-}$ & 0.322 & 5$^{-}$ & 0.730        \\
 0.079 & 0$^{-}$ &0.355 & 5$^{-}$ &  0.768    \\

0.106 & 2$^{-}$ &  0.378 & 4$^{+}$& 0.823         \\
 
 0.127 & 5$^{+}$ &0.381 & 1$^{-}$&  0.863        \\
 0.133 & 1$^{-}$ &0.421 & 2$^{-}$ & 0.914   \\

 0.139 & 1$^{+}$& 0.426 & 5$^{+}$& 0.976         \\
 0.139 & 2$^{-}$&0.478 & 1$^{+}$ &  1.002      \\
0.156 & 3$^{-}$&0.480 & 3$^{-}$& 1.052       \\
0.168 & 2$^{+}$&0.508 & 6$^{+}$&  1.086         \\
  0.177 & 5$^{-}$& 0.515 & 2$^{-}$&    1.129      \\
0.200 & 3$^{+}$& 0.520 & 2$^{+}$& 1.150    \\
0.223 & 0$^{+}$&0.523 & 6$^{-}$ &  1.198       \\
0.233 & 1$^{+}$ &   0.558 & 4$^{-}$          & 1.252  \\
0.237 & 3$^{-}$&  0.572 & 7$^{-}$& 1.280   \\
0.244 & 4$^{-}$& & & 1.294      \\
0.258 & 4$^{-}$& & & 1.346   \\
0.265 & 4$^{+}$& & & 1.397   \\

\noalign{\smallskip}\hline
\end{tabular}
\end{table}

Fig.~\ref{fig:3} shows the comparison of measured data with CRC calculations. As can be seen the inclusion of the 2$^{+}$ resonance state (E = 3.03~MeV) of $^8$Be  makes a large difference to $\sigma_{-1n}$. Recently, similar observation was made in $^9$Be+$^{197}$Au system~\cite{mk1}. Further, the effect of excited states of $^{160}$Tb is also illustrated in the figure. In the first case,  only excited states  with well determined J$^{\pi}$ values~\cite{nndc} are included (blue dashed line, E$_x<$ 0.572 MeV). The second set  was carried out by additionally including some  excited states between 0.572~MeV $< E_x <$ 1.397~MeV (see Table~\ref{tab:4}).  Since the J$^{\pi}$ information is not available for these states,  calculations are done with random J(1, 2) and $\pi (-,+)$ assignment (red dot-dash line).
It can be seen from the figure that both these  sets reproduce the observed trend of the measured cross section. The calculations with states upto E$_x$=0.572 MeV  under-predict the data, while those with states upto E$_x$ = 1.397~MeV are somewhat higher than the measured values.
The best agreement with data is observed with inclusion of states upto 
E$_x$ = 1.198 MeV (black solid line). 
It can be mentioned that calculations with an arbitrary assignment of fixed J$^{\pi}$ of 2$^+$ to  states at $E_x$ between 0.572 and 1.198~MeV, yield about 10$\%$ variation in the $\sigma_{CRC}$. Thus, the data is well described by the  CRC calculations including the  ground state and  the 2$^{+}$ resonance state of $^{8}$Be, and $^{160}$Tb excited states upto E$_x$=1.198 MeV.\\

\section{Systematics of 1$n-$ stripping cross section}

 For better understanding of systematics, a comparison of  $\sigma_{-1n}$ of different projectile-target combinations is generally presented in terms of   reduced energy (E$_{red}$) and cross sections ($\sigma_{red}$)  defined as~\cite{prsg}, 
\begin{equation}
    E_{red} = \frac{E_{cm}}{Z_{P}.Z_{T}/(A_{P}^{1/3}+A_{T}^{1/3})}  \\
   \end{equation}

\begin{equation}
       \sigma_{red} = \frac{\sigma_{-1n}}{(A_{P}^{1/3}+A_{T}^{1/3})^{2}}. \\
\end{equation}
where $Z_{P} (Z_{T})$ and $A_{P} (A_{T})$ refer to projectile (target) atomic number and mass number, respectively.

\begin{figure}

\resizebox{0.5\textwidth}{!}{%
  \includegraphics{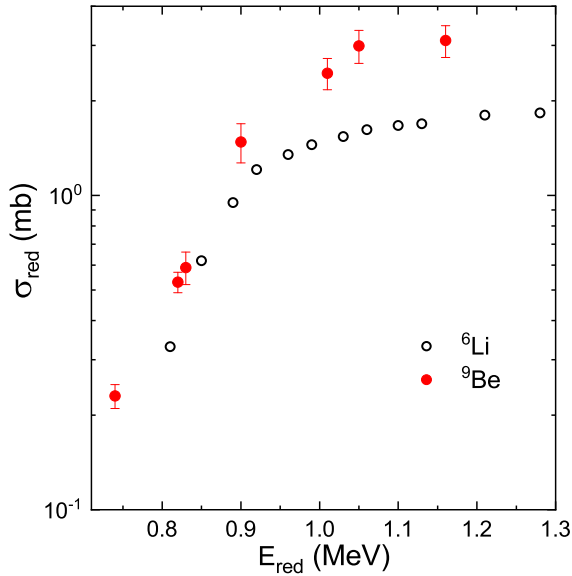}
}
\caption{Comparison of reduced 1$n$-stripping cross sections in  $^6$Li~\cite{mkp} and $^9$Be (present work) on $^{159}$Tb.}
\label{fig:4}       
\end{figure}

As mentioned earlier, 
the 1$n$-stripping measurements with $^{6}$Li on the $^{159}$Tb target are reported in Ref.~\cite{mkp}. However, $\sigma_{-1n}$ was deduced from an online measurement of the de-excitation of  63.68 keV $\gamma$-ray, and does not include the contribution from the transfer to the ground state. 
Figure~\ref{fig:4} shows reduced 1$n$-stripping  cross sections in  $^6$Li~\cite{mkp} and $^9$Be (present work) on $^{159}$Tb.

 It can be seen that  $\sigma_{red}$ for $^{9}$Be is higher as compared to that for $^{6}$Li, especially at higher energies. This is consistent with higher Q-value of $^9$Be (4.71 MeV) as compared to 0.71 MeV for $^6$Li. Similar enhancement was observed for $^{197}$Au target over energy range of 0.76$-$1.16 V$_b$~\cite{mk1}.

\begin{figure}[h]

\resizebox{0.5\textwidth}{!}{%
  \includegraphics{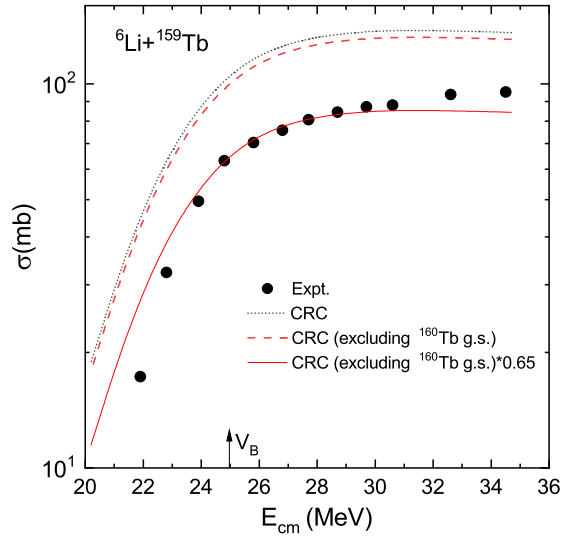}
}
\caption{ The $\sigma_{-1n}$(CRC)  in $^6$Li +$^{159}$Tb together with data from Ref.~\cite{mkp}. Excited states of $^{160}$Tb upto 0.73~MeV are considered in the calculation.  }
\label{fig:5}       
\end{figure}
In ref.~\cite{mkp}, the 1$n$-stripping cross section data was compared with the  DWBA calculations including only the first excited state of $^{160}$Tb at  63.68 keV. Although  a reasonable agreement was seen in the vicinity of the barrier with C$^{2}$S = 0.25,  the observed energy dependence was not well reproduced.
Hence, for better understanding, the CRC calculations are carried out in the $^6$Li+$^{159}$Tb, following the procedure described in the previous section. 
The global optical potential parameters for $^{6}$Li~+~$^{159}$Tb system and the binding potential parameters for $n$+$^{5}$Li are taken from Ref.~\cite{jc}, and  are listed in Table~\ref{tab:3}. The binding potential parameters for $n$~+ $^{159}$Tb are taken to be  same as in the $^{9}$Be~+~$^{159}$Tb calculations. The spectroscopic factor for $^{6}$Li/$^{5}$Li is taken to be 1.12~\cite{mbt}, while that for all target states is taken to be 1.0. The CRC calculations including the ground state of $^{5}$Li and the excited states of $^{160}$Tb upto 0.73 MeV (i.e. upto Q$_{gs}\sim0.71$MeV)
are shown in Fig.~\ref{fig:5} (dotted line). 
Since the data does not include the contribution of the ground state, calculations without the ground state of $^{160}$Tb are also shown in the same figure (red dashed line) and it is evident that contribution from the ground state is negligibly small. 
It should be pointed out that the observed trend is well reproduced by the present calculation over entire energy range, although calculations over-predict the data. From the figure, it can be seen that the CRC calculations with a scale factor of 0.65 (red solid line) well describe the data. The scale factor accounts for the deviation of  spectroscopic factors from the assumed value of 1 in the calculation.
Thus, the present CRC calculations are able to give a better overall description of the data. 

\begin{figure}

\resizebox{0.5\textwidth}{!}{%
  \includegraphics{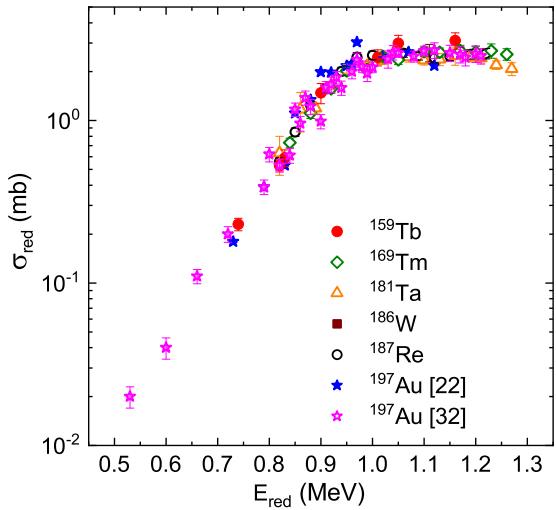}
}
\caption{Comparison of reduced 1$n$-stripping  cross sections in $^{9}$Be~+~X systems, where X represents $^{159}$Tb (present work), $^{169}$Tm~\cite{ydf}, $^{181}$Ta~\cite{ydf}, $^{186}$W~\cite{ydf1}, $^{187}$Re~\cite{ydf}, $^{197}$Au~\cite{mk1,fg}.}
\label{fig:6}       
\end{figure}
Finally, we present a systematic comparison of $\sigma^{red}$(-1n) of $^{9}$Be on various heavy target nuclei (A $>$ 150). Figure~\ref{fig:6} shows the $\sigma_{red}$(-1$n$) for  different targets -
$^{159}$Tb (present work), $^{169}$Tm~\cite{ydf}, $^{181}$Ta~\cite{ydf}, $^{186}$W~\cite{ydf1}, $^{187}$Re~\cite{ydf}, and $^{197}$Au~\cite{mk1,fg}. The ground state Q-values of 1$n$ - stripping reaction in these systems are in the range of $\sim$ 4-5 MeV and are given in Table~\ref{tab:5}. It is evident from the figure that no strong dependence on the target is observed. 

\begin{table}
\centering
\caption{Ground state Q-value for 1$n$-stripping in $^{9}$Be~+~$\rm X$ systems.}
\label{tab:5}       
\begin{tabular}{ll}
\hline\noalign{\smallskip}
System & Q-value (MeV)  \\
\noalign{\smallskip}\hline\noalign{\smallskip}
$^{9}$Be~+~$^{159}$Tb  & 4.71 \\
$^{9}$Be~+~$^{169}$Tm  & 4.93 \\
$^{9}$Be~+~$^{181}$Ta  & 4.39  \\
$^{9}$Be~+~$^{186}$W  & 3.80 \\
$^{9}$Be~+~$^{187}$Re  & 4.20 \\
$^{9}$Be~+~$^{197}$Au  & 4.85 \\

\noalign{\smallskip}\hline
\end{tabular}
\end{table}

\section{Summary and conclusions}
Neutron transfer reactions, especially in weakly bound nuclei,  are important to understand interplay of various reaction mechanisms. Measurements of $\sigma_{-1n}$ in $^{9}$Be + $^{159}$Tb system over a wide energy range,  E$_{lab}\sim$ 27 to 43 MeV are carried out using offline gamma counting method. The cross sections are extracted from the observed yield of 879 keV gamma ray, with appropriate decay corrections. The observed half-life of 879 keV gamma ray, 77$\pm$9 d, is in good agreement with the literature value of 72.3 d.  The  measured excitation function is well described by CRC calculations, including $^{8}$Be ground state and 2$^{+}$ resonance state along with $^{160}$Tb excited states up to 1.198~MeV.

The present $\sigma_{-1n}$ data of $^9$Be projectile is compared with existing data for $^6$Li on $^{159}$Tb and is found to be significantly higher.
A comparative study of $1n$-stripping cross section for $^{9}$Be projectile on different targets in A $\sim$ 150 region (Z $\sim$ 65-79) indicates that the cross section is nearly independent of the target nucleus.  
It will be interesting to study such systematics with other weakly bound unstable nuclei.

 \section{Acknowledgments}
We thank the PLF staff for providing the steady and smooth beam during the experiments and the target laboratory personnel for their help in the target preparation. We acknowledge the support of the Department of Atomic Energy, Government of India, under Project No. 12P-R$\&$D-TFR-5.02-0300.

%
%

\end{document}